\documentclass[twocolumn,showpacs,preprintnumbers,prc,superscriptaddress]{revtex4-2}
\usepackage{graphicx,color}
\usepackage{float}
\usepackage{bm}
\usepackage[pdftex,colorlinks=true, linkcolor = blue, citecolor=blue,urlcolor=blue, bookmarksnumbered=true, bookmarksopen=true]{hyperref}
\begin{document}

\title{Influence of non-statistical properties in nuclear structure 
       on emission of prompt fission neutrons}

\author{T. Kawano}
\email{kawano@lanl.gov}
\affiliation{Los Alamos National Laboratory, Los Alamos, NM 87545, USA}

\author{S. Okumura}
\email{S.Okumura@iaea.org}
\affiliation{NAPC--Nuclear Data Section, International Atomic Energy Agency, Vienna A-1400, Austria}

\author{A.~E. Lovell}
\affiliation{Los Alamos National Laboratory, Los Alamos, NM 87545, USA}

\author{I. Stetcu}
\affiliation{Los Alamos National Laboratory, Los Alamos, NM 87545, USA}

\author{P. Talou}
\affiliation{Los Alamos National Laboratory, Los Alamos, NM 87545, USA}

\date{\today}
\preprint{LA-UR-21-22687}

\begin{abstract}
The Hauser-Feshbach Fission Fragment Decay (HF$^3$D) model is extended
to calculate the prompt fission neutron spectrum (PFNS) for the thermal
neutron induced fission on $^{235}$U, where the evaporated neutrons
from all possible fission fragment pairs are aggregated. By studying
model parameter sensitivities on the calculated PFNS, as well as
non-statistical behavior of low-lying discrete level spin
distribution, we conclude that discrepancies between the aggregation
calculation and the experimental PFNS seen at higher neutron emission
energies can be attributed to both the primary fission fragment yield
distribution and the possible high spin states that are not predicted
by the statistical theory of nuclear structure.
\end{abstract}
\pacs{24.60.-k,24.60.Dr}
\maketitle

\section{Introduction}
\label{sec:introduction}

The nuclear fission process produces a few prompt fission neutrons and
a lot of $\gamma$-rays to release all the available excitation
energies. Generally we assume that the majority of these prompt
particles are emitted from two well-separated fission fragments after
scission. The neutrons that are evaporated from the two moving fission
fragments, which are assumed to be emitted isotropically in the
center-of-mass system (CMS) of each fragment, are boosted into the
laboratory frame (LAB)~\cite{Feather1942, Terrell1959}, and observed
as the prompt fission neutron spectrum (PFNS, $\chi(E)$).  When a
simple evaporation mechanism is assumed, $\epsilon\exp(-\epsilon/T)$,
where $\epsilon$ is the CMS neutron energy and $T$ is the temperature
parameter, the PFNS in the LAB frame is given by the
model proposed by Madland and Nix~\cite{Madland1981}. When a
statistical Hauser-Feshbach decay~\cite{Hauser1952} is performed from
individual fission fragment pairs~\cite{Browne1974, Litaize2010,
Becker2013, Regnier2016}, the aggregated neutron and $\gamma$-ray energy spectra
--- the Hauser-Feshbach Fission Fragment Decay (HF$^3$D) model ---
should represent direct connection between the statistical decay
inside a compound nucleus and experimentally observable
data~\cite{Okumura2018, Lovell2021, Okumura2021}.  Although such
approaches should include the most physics, a long-standing problem
remains in the calculated PFNS at higher outgoing neutron energies,
namely the energy spectrum on the high energy side (typically above
5~MeV) tends to fall more quickly than experimental
data~\cite{Becker2013}.

The problem is primarily due to the energy conservation.  An available
energy for the evaporating neutron is tightly confined by the fission
fragment kinetic energy, the neutron separation energy, and the
reaction $Q$-value for a fixed mass split. The Hauser-Feshbach neutron
spectrum will be strictly zero beyond the energetically allowed point,
in contrast to the simple evaporation spectrum that can go infinity.
Due to the requisite energy conservation, the HF$^3$D model often
gives too soft shape nevertheless the average energy in the LAB system
tends to be consistent with experimental data~\cite{Kawano2013}.  It
is also reported that the calculated spectrum-average cross sections
$\overline{\sigma} = \int \sigma(E) \chi(E) dE$ tend to be smaller
than some measured values especially for high threshold energy
reactions such as (n,2n)~\cite{Capote2016}.  This has been discussed
for a long time by studying model parameter sensitivities, such as the
optical model potential, level densities, and $\gamma$-ray strength
functions, nevertheless a clear explanation has not been made yet.

Many theoretical and experimental studies in the past have
confirmed that the fission process produces two fission fragments
where high-spin compound states are favorably
populated~\cite{Wilhelmy1972, Bonneau2007b, Stetcu2013, Bertsch2019,
Wilson2021}.  This implies that a compound nucleus decaying by
emitting a neutron might be enhanced if the low-lying levels have
higher spin $I$ nevertheless the statistical spin distribution predicts a
very small probability.  The similar phenomenon is reported to the
prompt fission $\gamma$-ray spectrum (PFGS) by Makii et
al.~\cite{Makii2019}. They argue ``non-statistical properties'' in
nuclear structure, which may enhance $\gamma$ decay from a high-spin
compound state directly to low-lying levels by the E1 transition, if
unknown discrete states have somewhat higher spin $I$ than the
statistical spin distribution predicts.

In the HF$^3$D model, we take nuclear structure information from the
RIPL-3 database~\cite{RIPL3}, which is based on the evaluated nuclear
structure data files, ENSDF~\cite{ENSDF}.  When applied to HF$^3$D
where we look for nuclear structure properties (excitation energy
$E_x$, spin $I$, and parity $\pi$) of many fission fragments, and this
information might be incomplete since many of the fragments are
neutron-rich and unstable.  To supplement the missing $I^\pi$ values,
we employ the level density formula~\cite{Gilbert1965}, which
implicitly imposes a well-behaved statistical distribution on the
low-lying levels. By considering the non-statistical argument in
PFGS~\cite{Makii2019}, we envision this could be one of the keys to
solve the PFNS issue as well. In this paper, we calculate the thermal
neutron induced fission on $^{235}$U with the HF$^3$D model, and
discuss the impact of the non-statistical properties in the nuclear
structure on the fission observables.

\section{Theory}
\label{sec:theory}

\subsection{Hauser-Feshbach fission fragment decay model}

The nuclear fission produces a pair of highly excited fission
fragments, and they de-excite by emitting several prompt neutrons and
$\gamma$-rays.  The Hauser-Feshbach fission fragment decay (HF$^3$D)
model calculates this de-excitation process by the statistical
Hauser-Feshbach theory, where the decay probabilities are governed by
the neutron and $\gamma$-ray transmission coefficients and the level
densities at the final states.  The neutron transmission coefficients
are given by solving the Schr\"{o}dinger equation for the complex
optical potential. We employ the global optical potential parameters
of Koning and Delaroche~\cite{Koning2003}. The Giant Dipole Resonance
(GDR) model with the GDR parameters reported in RIPL-3~\cite{RIPL3}
generates the $\gamma$-ray transmission coefficients. A
phenomenological model~\cite{Gilbert1965, Kawano2006} is employed to
calculate the nuclear level densities. At low excitation energies,
discrete levels given in RIPL-3 are explicitly included instead of the
level density model.  RIPL-3 often includes some discrete levels to
which the excitation energy $E_x$ is experimentally known but its spin
$I$ and/or parity $\pi$ are uncertain.  We assign $I^\pi$ of these
states by the Monte Carlo sampling from the spin and parity distributions
of
\begin{equation}
  R(I,\pi) = \frac{1}{2} \frac{I+1/2}{\sigma^2}
         \exp\left\{
               -\frac{(I+1/2)^2}{2\sigma^2} 
             \right\} \ ,
  \label{eq:rj}
\end{equation}
where $\sigma^2$ is the spin cut-off parameter
\begin{equation}
  \sigma^2(E_x) = 0.006945 \sqrt{\frac{U}{a}} A^{5/3} \quad \hbar^2 \ .
  \label{eq:sigma2}
\end{equation}
$U$ is the pairing energy ($\Delta$) corrected excitation energy $U =
E_x - \Delta$, $a$ is the energy-dependent level density parameter,
and $A$ is the mass number. A 50\%-50\% partition is assumed for the
parity distribution. In the low-excitation energy region,
$\sigma^2(E_x)$ is replaced by a constant $\sigma_0^2$ estimated from
discrete level data. In the fission product mass region, this is
approximated by
\begin{equation}
 \sigma_0^2 
 = - 3.04\times 10^{-4} A^2 + 0.141 A - 0.265 \quad \hbar^2 \ ,
 \label{eq:sigma0}
\end{equation}
and this is connected to Eq.~(\ref{eq:sigma2}) when 
$\sigma^2(E_x) > \sigma_0^2$.

We perform the statistical Hauser-Feshbach decay calculations for all
the produced fission fragments when their yield $Y_P(Z,A)$ is more
than $10^{-6}$. Typically there are more than 400 fission fragments
that are characterized by an initial configuration $P(J,\Pi,E_x)$ ---
distributions of spin, parity, and excitation energy.  The calculated
results are averaged by weighting the fission fragment mass and charge
distributions.  To perform the five-fold integration over the
distributions of $Y_P(Z,A)$ and $P(J,\Pi,E_x)$, the Monte Carlo
sampling technique has been utilized~\cite{Becker2013, Litaize2010,
  Talou2018}. In contrast to the stochastic method, HF$^3$D
deterministically performs the numerical integration over these
distributions, such that all of the contributing fission fragment
pairs are properly included to calculate the averaged quantities. More
details are given in our previous publications~\cite{Okumura2018,
  Lovell2021}. Note that there exist more statistical decay codes,
which are similar to our model but do not perform the Hauser-Feshbach
calculation, see e.g. Refs.~\cite{Talou2018, IAEA-NDS-230}.

\subsection{Prompt fission neutron spectrum}

In the PFNS model by Maldand and Nix~\cite{Madland1981}, either the
most probable partition or some representative pairs are
included~\cite{Ohsawa1999, Iwamoto2008}. Alternatively, an aggregation
calculation is proposed by Tudora and Hambsch~\cite{Tudora2017}.
HF$^3$D is also the aggregation calculation, but the Hauser-Feshbach
statistical decay ensures the spin and parity conservation at each of the
decay stages.

The CMS neutron spectra $\phi_{L,H}$ from light ($L$) and heavy ($H$)
fission fragments, which implicitly include all the sequence of 
multiple neutron emission ($A_{L,H} \to A_{L,H}-1 \to A_{L,H}-2 \ldots$),
are normalized to the neutron multiplicity
$\nu_{L,H}$,
\begin{eqnarray}
 \int dE_x\!\int d\epsilon &\sum&_{J\Pi}
 \phi_{L,H}(J,\Pi,E_x,\epsilon) P_{L,H}(J,\Pi,E_x) \nonumber \\
  &=& \nu_{L,H} \ ,
  \label{eq:phi}
\end{eqnarray}
where the initial configuration $P_{L,H}(J,\Pi,E_x)$ has the normalization
\begin{equation}
 \int\sum_{J\Pi} P_{L,H}(J,\Pi,E_x) dE_x = 1 \ .
\end{equation}

The total kinetic energy TKE is our model input, taken from experimental
data, and the fragment kinetic energies $T_{L,H}$ are determined
by a simple kinematics
\begin{equation}
 T_L = {\rm TKE} \frac{A_H}{A_L + A_H}, \qquad T_H = {\rm TKE} \frac{A_L}{A_L + A_H} \ .
 \label{eq:ek}
\end{equation}
The CMS neutron spectra of light and heavy fragments are transformed
into the LAB frame using Feather's formula~\cite{Feather1942,
  Terrell1959}
\begin{eqnarray}
 \psi_{L,H}(E)
   &=& \int dE_x \!
       \int_{(\sqrt{E} - \sqrt{T_{L,H}})^2}^{(\sqrt{E} + \sqrt{T_{L,H}})^2} d\epsilon
       \frac{1}{4\sqrt{T_{L,H} \epsilon}} \nonumber\\
   &~& \sum_{J\Pi} \phi_{L,H}(J,\Pi,E_x,\epsilon) P_{L,H}(J,\Pi,E_x) \ .
 \label{eq:psi}
\end{eqnarray}
$\psi_{L,H}(E)$ carries the same normalization as Eq.~(\ref{eq:phi}),
and PFNS is given by averaging $\psi_{L,H}^{(k)}(E)$ weighted by 
the fission fragment yield $y_k$, where $k$ is an index for 
a particular $L$ and $H$ pair
\begin{equation}
 \overline{\psi}(E)
  = \sum_k \left\{ \psi_L^{(k)}(E) + \psi_H^{(k)}(E) \right\}y_k \ ,
 \label{eq:totalpsi}
\end{equation}
which is normalized to the average number of prompt fission neutron
multiplicity $\overline{\nu}$. Finally, the normalized PFNS is given by
\begin{equation}
 \chi(E) = \frac{\overline{\psi}(E)}{\overline{\nu}}  \ .
\end{equation}

Since the HF$^3$D model was originally developed to calculate the
independent and cumulative fission product yields, a very fast
algorithm to integrate the excitation energy distribution in
Eq.~(\ref{eq:phi}) was performed. However, this is not necessarily the best approach for
the PFNS calculation, because $T_{L,H}$ depends on the average
excitation energy of each fragment, $E_{x,L}$ and $E_{x,H}$. In the
PFNS calculation we discretize $P_{L,H}(J,\Pi,E_x)$ into excitation
energy bins, and perform the CMS-LAB conversion at each $E_x$. Detail
of the algorithm is given in Appendix~\ref{sec:appendix}. In this
technique the CMS-LAB conversion is only exact for the first neutron
emission, since we ignore a recoil effect.  Even though the multiple
neutron emission is treated approximately, this is negligible since
typical fission fragment masses are 80 or heavier compared to the
neutron mass.  Because of numerical integration, the computational
time strongly depends on the discretization bin width, and the lowest
emission energy is limited to the energy bin size (typically
100~keV). This is in contrast with the Monte Carlo technique, where
the emitted neutron energy can be arbitrary~\cite{Talou2018}.

\section{Result and Discussion}
\label{sec:result}

\subsection{PFNS by the HF$^3$D model}

The HF$^3$D model parameters are taken from our previous
works~\cite{Okumura2018, Okumura2021} for the thermal neutron induced
fission on $^{235}$U. The model parameters include the primary fission
fragment mass distribution $Y_P(A)$, TKE, the anisothermal parameter
$R_T$~\cite{Ohsawa1999, Ohsawa2000}, a scaling factor $f_J$ applied to
the spin distribution of initial configuration, and scaling parameters
$f_Z$ and $f_N$ in the odd-even staggering in the charge
distribution~\cite{Wahl1988, LA13928}. $Y_P(A)$ is represented by a
few Gaussians, so that the Gaussian width $\sigma_G$, the mass shift
measured from the symmetric fission $\Delta_G$, and the fraction of
each Gaussian component $F_G$ characterize $Y_P(A)$.

We take two sets of the parameters. The first set (prior) included
minimum effort of parameter optimization~\cite{Okumura2018}, and the
second set (posterior) was obtained by adjusting to experimental
fission yield data as well as average number of prompt and delayed
neutrons~\cite{Okumura2021}.

\begin{figure}
  \begin{center}
    \resizebox{\columnwidth}{!}{\includegraphics{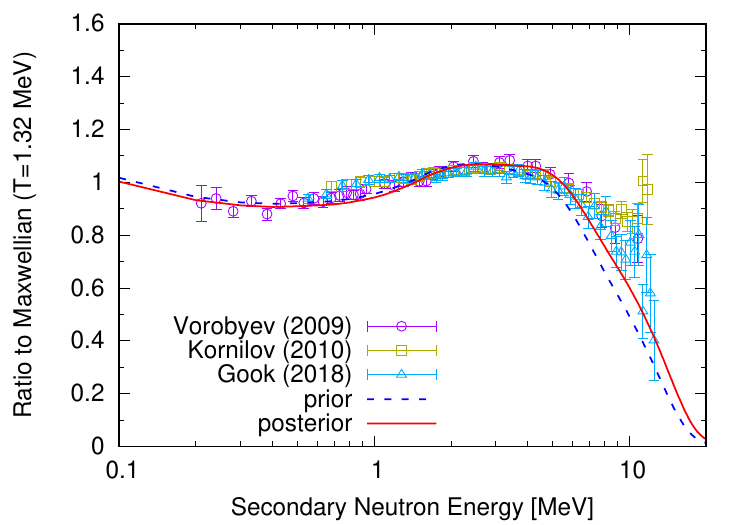}}
  \end{center}
  \caption{Comparisons of PFNS predicted by the HF$^3$D model with 
   experimental data. PFNS is shown by ratios to the Maxwellian at 
   the temperature of $T = 1.32$~MeV.}
  \label{fig:maxratio}
\end{figure}

Figure~\ref{fig:maxratio} shows the calculated PFNS as ratios to
Maxwellian
\begin{equation}
   \psi_M(E) = \frac{2}{\sqrt{\pi T^3}} \sqrt{E}
                      \exp\left( - \frac{E}{T} \right),
   \label{eq:maxwellian}
\end{equation}
at the temperature $T$ of 1.32~MeV, and compared with experimental
data~\cite{Vorobyev2009, Kornilov2010, Gook2018}. The average energies
of these spectra are 1.95~MeV for the prior set and 1.99 for the
posterior set.  Although the adjustment procedure does not include
these PFNS data, the posterior set better reproduces the experimental
data up to 7~MeV or so. However, the calculated PFNS still
underestimates data at higher outgoing energies.

\subsection{Partial neutron spectrum from fission fragment pairs}

\begin{figure}
  \begin{center}
    \resizebox{\columnwidth}{!}{\includegraphics{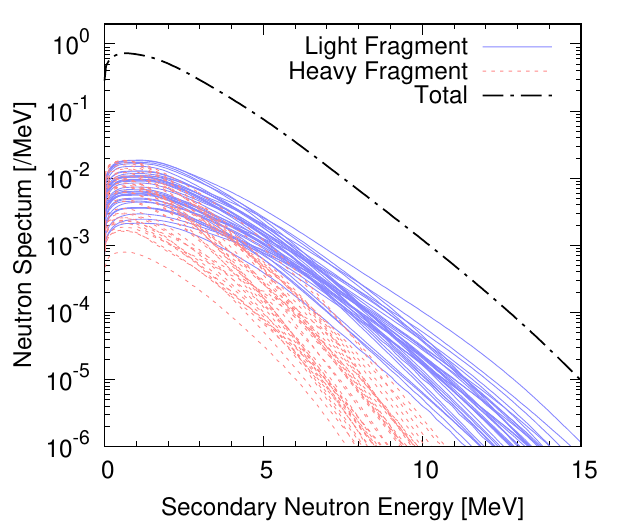}}
  \end{center}
  \caption{Individual contribution from each fission
    fragment to PFNS only for the
    fragment yield larger than 1\%.  The light solid curves are from
    the light fragments, and the dotted curves are from the heavy
    fragments.}
  \label{fig:U235pri_indiv}
\end{figure}

\begin{figure}
  \begin{center}
    \resizebox{\columnwidth}{!}{\includegraphics{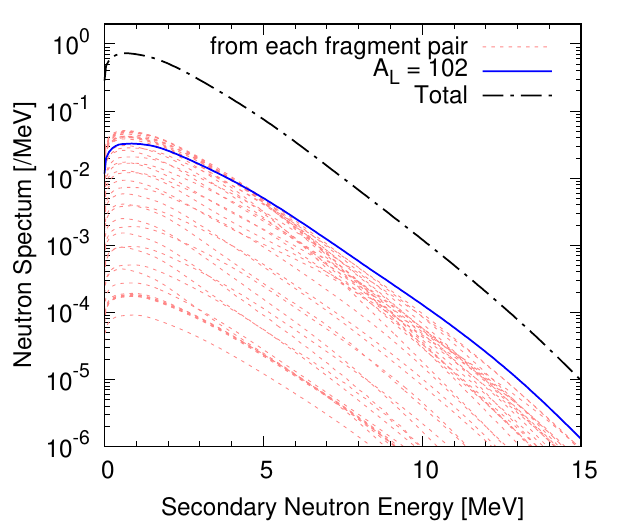}}
  \end{center}
  \caption{Partial contributions from each of fission fragment pair to
    PFNS, where all the same mass
    splits are lumped.}
  \label{fig:U235pri_mass}
\end{figure}

\begin{figure}
  \begin{center}
    \resizebox{\columnwidth}{!}{\includegraphics{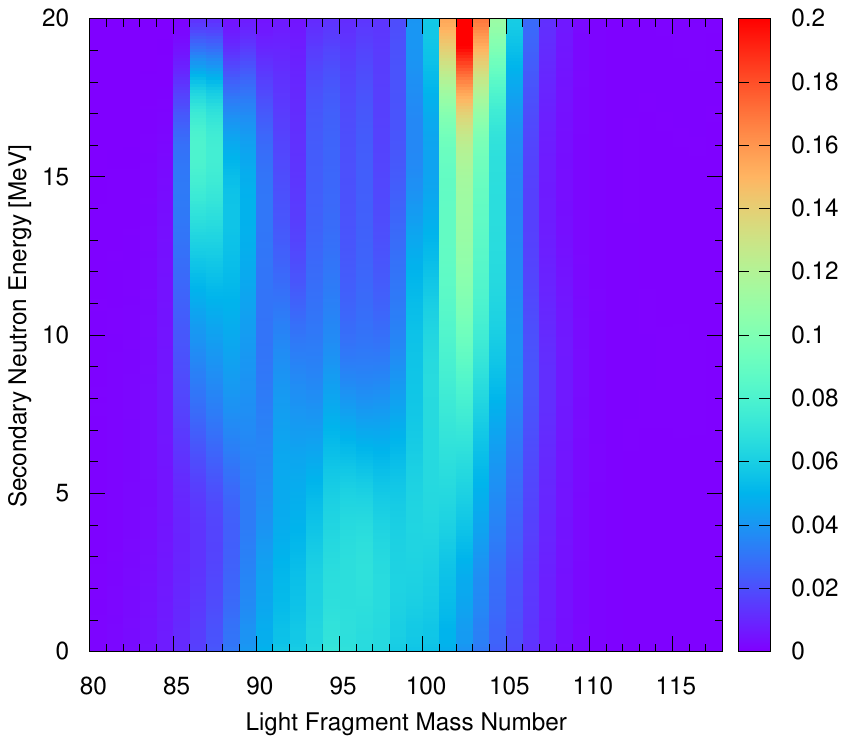}}
  \end{center}
  \caption{Partial fission spectra for a fixed $A$ shown as ratios
   to the total neutron emission, $\psi(A_L,E) / \overline{\psi}(E)$.}
  \label{fig:U235pri_mass3D}
\end{figure}

Aggregating neutron emission from individual fission fragment pair
gives us an insight into which mass partitions give a dominant
contribution to PFNS. To study which fission fragment pairs are
predominantly causing the difference between the prior and posterior
sets in Fig.~\ref{fig:maxratio}, we decompose the total neutron
emission $\overline{\psi}(E)$ for the prior set into each of
individual mass split $y_k \psi_{L,H}(E)$, which is shown in
Fig.~\ref{fig:U235pri_indiv}.  The fission pairs that have the yield
$y_k$ of larger than 1\% are plotted. The high energy tail in the
spectrum comprises of the neutrons from light fragments, because the
neutron separation energy, the neutron multiplicity, and fragment
kinetic energy are larger.  For better visibility, we lump
$\psi_{L,H}(E)$ for a fixed mass number $A$
\begin{equation}
 \psi(A_L,E)
 = \sum_k \left\{ \psi_L(E) + \psi_H(E) \right\}y_k \delta_{A,A_L} \ ,
 \label{eq:apsi}
\end{equation}
and this is shown in Fig.~\ref{fig:U235pri_mass}. The solid curve,
which is for $A_L=102$ and $A_H=134$, is the highest contribution to
the total spectrum above 5~MeV. We noticed the fission fragments near
these masses tend to give the harder spectrum tail, and this is
clearly shown by the ratio $\psi(A_L,E) / \overline{\psi}(E)$ in
Fig.~\ref{fig:U235pri_mass3D}. At lower outgoing energies,
contributions from each of fission fragment do not show particular
characteristics, and they basically follow the fission fragment
yield. However, at higher outgoing energies, some mass regions
($A\sim87$ and 102) have larger impact on lifting the PFNS tail.

\begin{figure}
  \begin{center}
    \resizebox{\columnwidth}{!}{\includegraphics{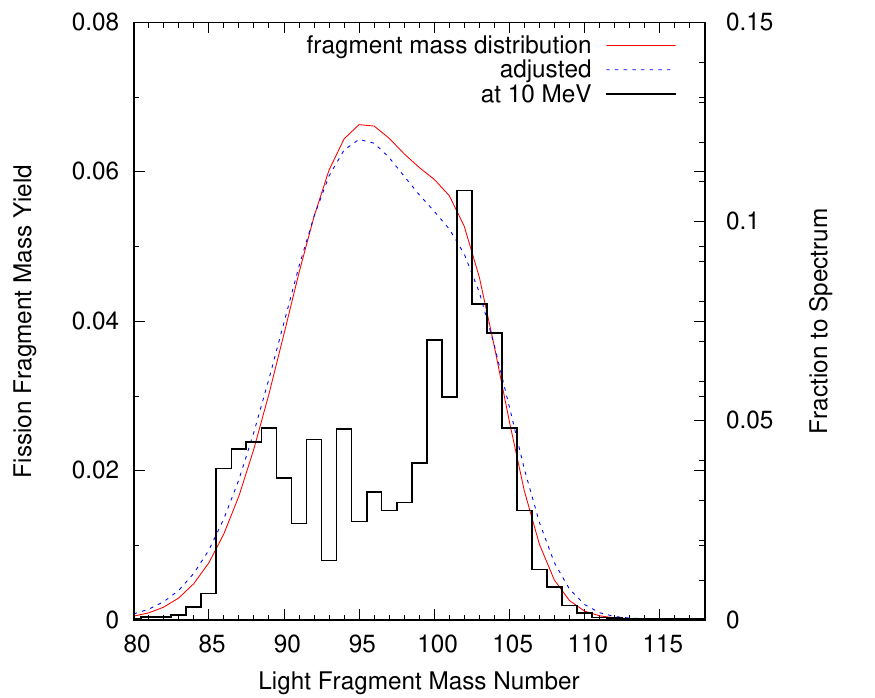}}
  \end{center}
  \caption{$\psi(A_L,E) / \overline{\psi}(E)$ for $E=10$~MeV (right axis),
    compared with the fission fragment distribution (left
    axis). Only the light fragment peak is shown.}
  \label{fig:fragdist}
\end{figure}

We can see this more clearly by slicing this distribution at the
outgoing energy of 10~MeV, and compare with the fission fragment
distribution, which is shown in Fig.~\ref{fig:fragdist}. The posterior
$Y_P(A)$, which was obtained by adjusting to some selected fission
product yield data, indeed becomes slightly wider distributions.
Unexpectedly the fragment yield at $A_L=102$ decreased after the
parameter tuning, and this reduction was compensated by the increase
in the outer ($A_L < 90$) and symmetric ($A_L > 102$) regions. This
indicates the wider fragment distribution may harden PFNS to some
extent, but restricted by reduction in the most sensitive component of
$A_L=102$.  We also studied other parameters, $R_T$, $f_J$, $f_Z$, and
$f_N$, as well as TKE. However, they have rather modest sensitivities
to PFNS, especially to its shape.

\subsection{Non-statistical properties in discrete levels}

The spin distribution in Eq.~(\ref{eq:rj}) was derived by an
assumption that the distribution of magnetic quantum number $m$ for
many particle many hole configurations forms Gaussian with the average
of zero~\cite{Gilbert1965}. It does not consider any particular
properties of nuclear states, such as the collective
excitation. Deformed even-even nuclei often show a rotational spectrum
($0^+$, $2^+$, $4^+, \ldots$), and sometimes we find higher $I$ states
at relatively low excitation energies, in contrast to a prediction by
the statistical theory. For example, a neutron-rich nucleus $^{102}$Zr
has very deformed shape ($\beta_2 = 0.36$)~\cite{Moller1995} and
exhibits the ground state rotational band structure at low energies.
However, experimentally known levels in the band are up to the 964-keV
($6^+$) level.

Because $\sigma^2$ as estimated in Eqs.~(\ref{eq:sigma2}) and
(\ref{eq:sigma0}) could have a relatively large uncertainty, and there
is no simple way to estimate how the actual spin distribution for a
specific nuclide deviates from the statistical prediction
unfortunately, we study some overall influence of the non-statistical
properties in the low-lying states by adjusting the spin cutoff
parameter $\sigma^2$.  We conduct this in two approaches; (i) modify
unknown $I$'s that are estimated by $\sigma^2$, and (ii) modify the
$I$-distribution of level density at low excitation energies.

\subsubsection{Unknown spin of discrete states}

\begin{figure}
  \begin{center}
    \resizebox{\columnwidth}{!}{\includegraphics{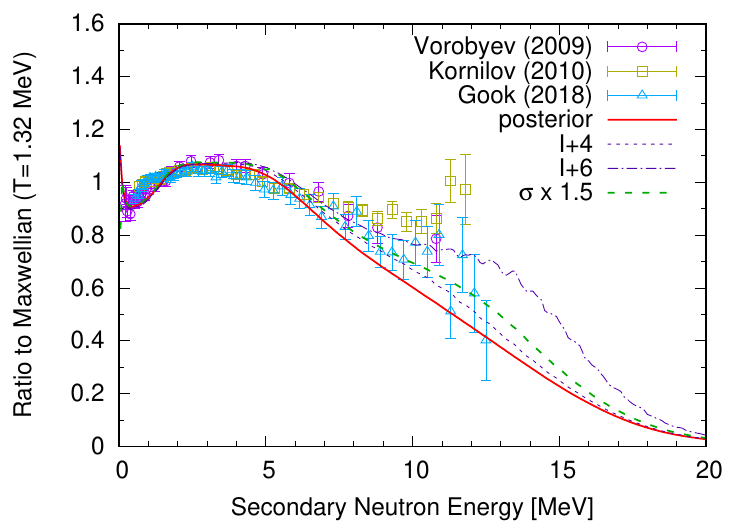}}
  \end{center}
  \caption{Comparisons of PFNS predicted by the HF$^3$D model with 
   experimental data, which is shown by ratios to Maxwellian
   the temperature of $T = 1.32$~MeV.
   When discrete level spins are unknown,
   they are increased by 4$\hbar$ and 6$\hbar$, or sampled from
   the spin distribution formula where $\sigma$ is multiplied by 1.5.}
  \label{fig:maxratio_spin}
\end{figure}

We assign a randomly sampled spin to the discrete level whenever $I$
is experimentally unknown (while its excitation energy is known).  The
sampling is based on Eq.~(\ref{eq:rj}), which implies the
$I$-distribution is statistical. To introduce non-statistical
fluctuation, we add 4$\hbar$ or 6$\hbar$ to the assigned spin.  Note
that this does not change any known $I$ of discrete levels, but only
modifies their unknown spin assignment.  The result is shown in
Fig.~\ref{fig:maxratio_spin}.  We see a significant increase in
the high energy tail of PFNS, assuming a non-statistical feature in
the discrete levels might be realistic. Although such the non-statistical
property may vary nucleus by nucleus, additional 4 -- 6$\hbar$ to the
unknown spins may largely reconcile the discrepancy between the
calculated PFNS and experimental data above 7~MeV.  These levels relax
strong hindrance of high-energy neutron emission from the compound
states that also have high spins.

The similar enhancement can be achieved by increasing $\sigma$ of
Eq.~(\ref{eq:rj}), and Fig.~\ref{fig:maxratio_spin} shows the case
when $\sigma$ is multiplied by 1.5. In this example, we found that
$^{94}$Rb is one of the dominant contributers causing the bump in the
10 -- 15~MeV region. However, this does not suggest any
missing high spin states in $^{94}$Rb, but just one of many
possibilities.

While these modifications changed the PFNS tail notably, increase in
the calculated prompt neutron multiplicity $\overline{\nu}$ was less
than 0.3\%.

\subsubsection{Spin distribution of level density at low excitation energies}

\begin{figure}
  \begin{center}
    \resizebox{\columnwidth}{!}{\includegraphics{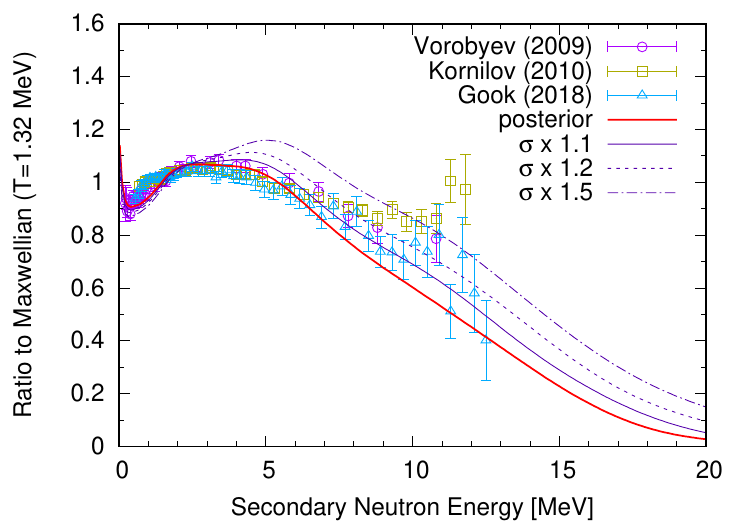}}
  \end{center}
  \caption{Comparisons of PFNS predicted by the HF$^3$D model with 
   experimental data. The spin cutoff parameter in the level density 
   formula at low excitation energies is multiplied by 1.1, 1.2, and 1.5.}
  \label{fig:maxratio_sigma}
\end{figure}

From Eq.~(\ref{eq:rj}) the average spin $\langle I \rangle$ is roughly
proportional to $\sigma$ ($\langle I \rangle \simeq 1.2\sigma$), and
typical values of $\sigma^2$ in the fission product mass region are 10
($\sigma \sim 3$) or so at low excitation energies; see
Eq.~(\ref{eq:sigma0}). When $\sigma$ is larger than these average
values, this also enhances neutron emission from the initially
populated compound state. Figure~\ref{fig:maxratio_sigma} shows how
PFNS changes its shape when $\sigma_0$ of Eq.~(\ref{eq:sigma0}) is
multiplied by 1.1, 1.2, and 1.5. In this case, all of the fission
fragments contribute to the shape-change, hence it tends to overshoot
the entire experimental data points near 6~MeV, albeit increase in
$\sigma$ lifts the PFNS tail. This is in contrast to
Fig.~\ref{fig:maxratio_spin}, where we increased $I$ only when it is
not assigned. This confirms that the deficiency in the model does not
reside in the statistically averaged parameters such as the estimated
$\sigma^2$ but in the incomplete nuclear structure information for
some specific nuclei.

The both cases shown above demonstrate that the non-statistical
feature in the spin distribution at low excitation energies is
sensitive to the tail of PFNS. By following the statistical theory for
the level spectrum, extremely high spin states are seldom found at low
excitation energies, since the small number of particle-hole
configurations hardly couple to high-$I$. However, the ground state
rotational band of even-even nucleus might exhibit $10\hbar$ or higher
levels, even though they are not easily observed experimentally.  Small
non-statistical contributions from each fission fragment would build
up in the aggregation calculation, and they finally impact on the PFNS
shape. Having said that, it should be noted that the high energy part of
PFNS in Figs.~\ref{fig:maxratio_spin} and
\ref{fig:maxratio_sigma} is exaggeratedly plotted by ratios to
Maxwellian, where the absolute PFNS varies in several orders of
magnitude in the MeV region. When we calculate spectrum average cross
sections $\overline{\sigma}$ for the high $Q$-value reactions, such as
the (n,2n) reaction, these values would be strongly influenced by
assumptions made for uncertain spin of low-lying states. Precise
spectroscopic studies on nuclear structure for neutron-rich nuclei are
essential to better understand the tail of PFNS.

\section{Conclusion}
\label{sec:conclusion}

We extended the Hauser-Feshbach Fission Fragment Decay (HF$^3$D) model 
to calculate the prompt fission neutron spectrum (PFNS).
We carefully investigated the long-standing problem of ``too soft
PFNS'' when the aggregated calculation of Hauser-Feshbach neutron
spectrum is performed for all the possible fission fragments, and we
found two important components in the HF$^3$D model. The primary
fission fragment distribution $Y_P(A)$, which is parameterized by
several Gaussians, needs to be slightly wider than experimentally
reported data. The wider distribution is also supported by other
fission observables, such as the cumulative fission product yields,
the prompt neutron multiplicity $\overline{\nu}_p$, and the
$\beta$-delayed $\overline{\nu}_d$. We demonstrated that the primary
fragment distribution accounts for the underestimation of PFNS up to
7~MeV. At higher outgoing energies, a possible reason of neutron
emission hindrance is a non-statistical nature of the spin
distribution in low-lying levels. When some discrete levels to which
the spin $I$ is not reliably determined, can have higher $I$ than
predicted spin by the statistical model, a high $J$ compound state is
able to decay directly to these state by emitting a neutron. This decay
process creates the higher energy tail in PFNS, and it may reconcile
discrepancies seen in the calculated and experimental spectrum average
cross sections.

\section*{Acknowledgment}

TK thanks Dr.~Nishio and Dr.~Makii of JAEA for valuable discussions.
This work was partially support by both the Nuclear Criticality Safety
Program and the Office of Defense Nuclear Nonproliferation Research \&
Development (DNN R\&D), National Nuclear Security Administration,
U.S. Department of Energy.  This work was carried out under the
auspices of the National Nuclear Security Administration of the
U.S. Department of Energy at Los Alamos National Laboratory under
Contract No. 89233218CNA000001.

%

\appendix*
\section{CMS-LAB conversion in the HF$^3$D model}
\label{sec:appendix}

The HF$^3$D model defines an initial population $P_{L,H}(E_x,J,\Pi)$
in a discretized compound nucleus (CN) with a given energy bin $\Delta E_x$,
which is a joint distribution of excitation energy $E_x$ and spin
$J$. The population is normalized as
\begin{eqnarray}
 &~& \sum_{J\Pi} \int P_{L,H}(E_x,J,\Pi) dE_x \nonumber \\
 &=& \sum_{J\Pi} \frac{1}{2} \int G_{L,H}(E_x) R(J) dE_x \nonumber \\
 &\simeq& \sum_{J\Pi} \frac{1}{2} R(J) \sum_k G_{L,H}^{(k)} \Delta E_x = 1 \ ,
\end{eqnarray}
where the excitation energy distributions, $G_{L,H}(E_x)$, for both the light
and heavy fragments are represented by a Gaussian
\begin{equation}
  G_{L,H}(E_x) = \frac{1}{\sqrt{2\pi} \delta_{L,H}}
            \exp\left\{-\frac{(E_x-E_{L,H})^2}{2\delta_{L,H}^2}\right\} \ ,
\end{equation}
with the average excitation energy of $E_{L,H}$ and the width of
$\delta_{L,H}$. The distribution of spin $J$ is assumed to be
proportional to the spin distribution in a level density
formula~\cite{Kawano2013},
\begin{equation}
  R_{L,H}(J) = \frac{J+1/2}{f_J^2\sigma_{L,H}^2(U)}
              \exp\left\{-\frac{(J+1/2)^2}{2 f_J^2\sigma_{L,H}^2(U)} \right\} \ ,
  \label{eq:rjp}
\end{equation}
where $\sigma^2(U)$ is the spin cut-off parameter, $U$ is the
excitation energy corrected by the pairing energy, and $f_J$ is an
adjustable scaling factor. This satisfies the normalization condition
of $\sum_{J} R(J) = 1$.  The parity distribution is just 1/2.

\begin{figure*}
 \begin{center}
  \begin{tabular}{cc}
    (a) & (b)\\
    \resizebox{0.8\columnwidth}{!}{\includegraphics{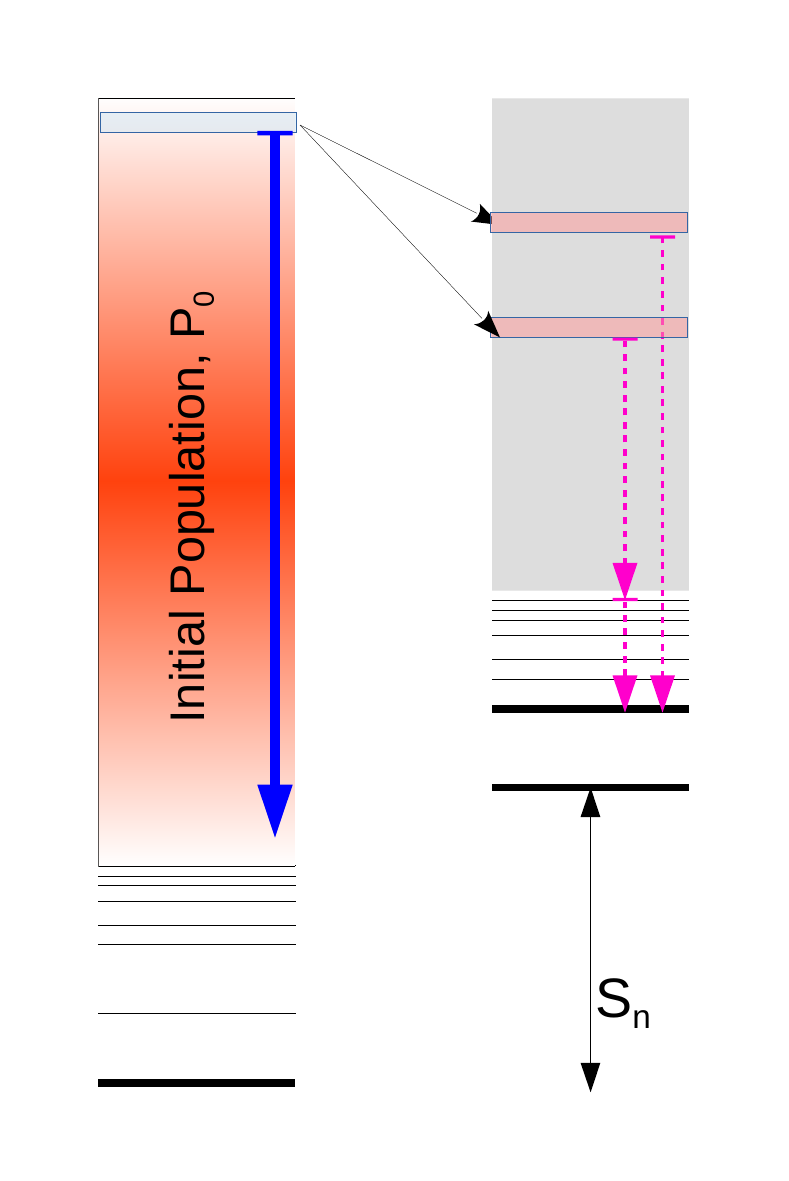}} &
    \resizebox{0.8\columnwidth}{!}{\includegraphics{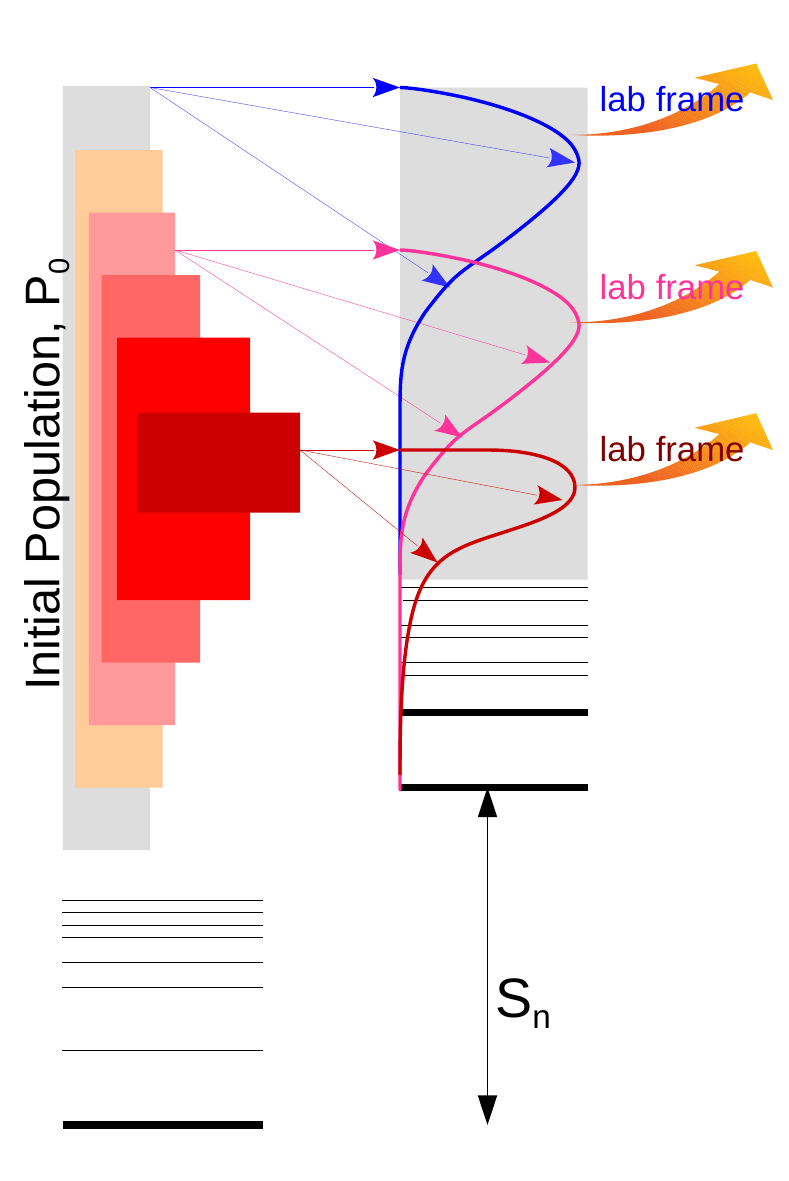}} \\
  \end{tabular}
 \end{center}
 \caption{Scheme of compound decay calculation for (a) the ground state production,
   and (b) prompt fission neutron spectrum.}
  \label{fig:CNdecaycalc}
\end{figure*}

Once $P_{L,H}(E_x,J,\Pi)$ is created in a CN, we start a statistical
decay calculation (neutron and $\gamma$-ray emission by the
Hauser-Feshbach theory) from the highest energy all way down to the
ground states of all the residual nuclei. Since $P_{L,H}(E_x,J,\Pi)$
is normalized to unity, the sum of the production probabilities of all
the residual nuclei also satisfies the normalization. This is
equivalent to integrate the initial population over the excitation
energy, spin, and parity. The resulting populations of each ground
state (or isomeric state if it does not further decay) equal to the
independent fission product yield. This is schematically shown in
Fig.~\ref{fig:CNdecaycalc} (a). Although this method simplifies the
three-fold integration of $E_x$, $J$, and $\Pi$, this method has a
problem when we apply it to the neutron spectrum in the laboratory
system, because the information of fragment kinetic energy is
excluded.

When we look into the energy distribution of the prompt fission
neutrons, we have to keep track of two fission fragments that are
flying away at a given kinetic energy, and the kinetic energy depends
on $E_x$. To do this, as in (b) of Fig.~\ref{fig:CNdecaycalc}, we
discretize the excitation energy (the energy-bin is typically 1~MeV,
which is wider than $\Delta E_x$ of 100~keV or so), and perform the
CMS-LAB conversion at each discretized energy-bin so that the kinetic
boost is properly included.  Although this algorithm yields exactly
the same ground state production probabilities as in the case (a), the
computational time will be significantly longer.

As a fast algorithm, the method (a) may still be useful for
calculating an approximated PFNS. First we calculate the average CMS
neutron spectrum, then transform it into the LAB frame by applying the
average fragment kinetic energies. However, this procedure tends to
yield a harder spectrum and the average energy would be higher by a
few percent.

\end{document}